\documentclass[twocolumn,showpacs,prl,amsmath,amssymb,superscriptaddress]{revtex4}

\usepackage{graphicx}
\usepackage{bm}
\usepackage{bbm}
\usepackage{mathbbol}

\DeclareMathAlphabet{\mathpzc}{OT1}{pzc}{m}{it}

\begin{document}

\title{Dephasing of two-spin states by electron-phonon interaction in
semiconductor nanostructures: Spin-boson model with a dissipative reservoir}
\author{Xuedong \surname{Hu}}
\affiliation{Department of Physics, University at Buffalo, SUNY, Buffalo, NY 14260-1500} 
\affiliation{Joint Quantum Institute, Department of Physics, University of Maryland, 
College Park, MD 20742} 
\date{\today}
\begin{abstract}
We study electron-phonon interaction induced decoherence between two-electron
singlet and triplet states in a semiconductor double quantum dot using a 
spin-boson model.  We investigate the onset and time evolution of this
dephasing, and study its dependence on quantum dot parameters such as dot
size and double dot separations, as well as the host materials (GaAs and Si).
We find that electron-phonon interaction causes an initial Gaussian decay
of the off-diagonal density matrix element in the singlet-triplet Hilbert space, 
and a long-time exponential decay originating from phonon relaxation.  
\end{abstract}

\pacs{
03.67.-a; 76.60.Lz; 03.65.Yz;
76.30.-v; 03.67.Lx; 76.90.+d}

\maketitle

In the past few years, significant experimental progresses in the study of
semiconductor spin qubits \cite{Petta,Koppens,Shaji,Lansbergen,Tarucha} have 
reconfirmed electron spins as one of the leading candidates for the building 
block of a solid state quantum information processor.  Recent theoretical
studies have also clarified single spin decoherence channels and their relative
importance in semiconductor quantum dots 
\cite{Golovach,deSousa,Witzel,Yao,Deng,Cywinski}, with hyperfine interaction
with environmental nuclear spins identified as the main culprit for electron
spin decoherence.  As experimental studies shift toward controlled coupling 
of spin qubits, quantum coherence properties of multiple-spin states are
naturally the next important theoretical problem, whose solution could go a 
long way in determining whether spin qubits in semiconductor nanostructures 
indeed have desired scalability for a practical quantum computer.

Decoherence of two-spin states in a coupled double quantum dot is crucial
to the operation of an exchange-based spin quantum computer architecture 
\cite{LD}.  Since nuclear spins are the main source
of single spin decoherence in quantum dots, existing studies have focused
on the decohering effects of the nuclear spins \cite{Coish,Yang}.  In addition, 
since exchange coupling is electrostatic
in nature, exchange-coupled electrons are vulnerable to charge noise
and other orbital fluctuations that have an electrical signature. 
For example, we have shown \cite{HD_PRL} that charge fluctuations lead to pure
dephasing by introducing noise into interdot barrier and/or interdot voltage 
bias.  

Here we study decoherence effects of electron-phonon
interaction on two-spin states in semiconductor nanostructures, based on
the consideration that lattice vibrations produce local electrical 
polarizations and could therefore affect the two-spin singlet and triplet
states (two-spin eigenstates in the absence of spin-orbit interaction) for 
exchange-coupled electrons.  We focus
on symmetrically coupled (i.e. no inter-dot voltage bias) quantum dots in GaAs 
and Si, and P donors in Si, all regarded as promising
candidates for qubits in spin-based quantum information processing.  Electron-phonon 
interaction is not spin-dependent and does not lead to direct 
transitions between singlet and triplet states.  Consequently
there is no phonon-induced relaxation between the singlet and
triplet states in the absence of spin-orbit interaction.  Our focus is
on pure dephasing effect of electron-phonon interaction
in a two-electron double dot.  Specifically, we obtain the effective interaction 
Hamiltonian, identify the most important types of
electron-phonon interaction, clarify
the dynamics of dephasing, and quantify its time scale
in GaAs and Si.  

The general electron-phonon interaction Hamiltonian in a semiconductor
takes the form \cite{YuCardona}
\begin{equation}
H_{ep} = \sum_{{\bf q},\lambda} M_{\lambda}({\bf q}) \rho({\bf q}) (a_{{\bf
q},\lambda} + a_{-{\bf q},\lambda}^\dagger) \,,
\end{equation}
where $a_{{\bf
q},\lambda}$ and $a_{-{\bf q},\lambda}^\dagger$ are phonon annihilation and 
creation operators with lattice momentum ${\bf q}$ and branch index $\lambda$,
and $\rho({\bf q})$ is the electron density operator.
For an electron to act as a spin qubit, its orbital degree of freedom needs to
be frozen in the ground state.  When two spin qubits are exchange-coupled in
a symmetric double quantum dot, their
orbital states are symmetric or anti-symmetric if their spin state is singlet
($|\!\!\uparrow\downarrow - \downarrow\uparrow\rangle/\sqrt{2}$) or triplet 
($|\!\!\uparrow\downarrow - \downarrow\uparrow\rangle/\sqrt{2}$, 
$|\!\!\uparrow \uparrow \rangle$, $|\!\!\downarrow \downarrow \rangle$).  Within the 
Heitler-London approximation, the two spatial wave functions can be written as
\begin{eqnarray}
\psi_S & = & \frac{1}{\sqrt{2(1+S^2)}} |L(1)R(2)+R(1)L(2) \rangle \,, \nonumber \\
\psi_{AS} & = & \frac{1}{\sqrt{2(1-S^2)}} |L(1)R(2)-R(1)L(2) \rangle \,, 
\end{eqnarray}
where $L$ and $R$ refer to the ground single-electron orbital states in the two 
dots, $S=\langle L|R\rangle$ is the overlap integral, and $1$ and $2$ are 
indices for the two electrons.  We can now project the electron-phonon interaction 
into the singlet-triplet Hilbert space.  All three triplet states have the same 
orbital wave function and cannot be differentiated by electron-phonon interaction.
The Hilbert space of interest is thus only two-dimensional, with the corresponding
basis states $\frac{1}{\sqrt{2(1+S^2)}} |L(1)R(2)+R(1)L(2)\rangle \times \frac{1}{\sqrt{2}}
|\!\uparrow\downarrow - \downarrow\uparrow\rangle$ and $\frac{1}{\sqrt{2(1-S^2)}} 
|L(1)R(2)-R(1)L(2)\rangle \times \frac{1}{\sqrt{2}}
|\!\uparrow\downarrow + \downarrow\uparrow\rangle$.  Since the Hamiltonian has 
no spin-dependence, the $2 \times 2$ electron-phonon
interaction Hamiltonian is diagonal:  
\begin{equation}
H_{eff} = \sum_{{\bf q},\lambda} M_{\lambda}({\bf q}) A_{\phi} \sigma_z
(a_{{\bf q},\lambda} + a_{-{\bf q},\lambda}^\dagger)\,,
\label{eq:Hamiltonian_eff}
\end{equation}
where $\sigma_z$ is a Pauli matrix in the two-dimensional two-electron Hilbert
space (it is not for single electron spins), and the charge distribution 
difference $A_\phi$ is given by
\begin{equation}
A_\phi = \frac{1}{2}[ \langle \psi_{AS}| \rho({\bf q}) |\psi_{AS} \rangle
- \langle \psi_{S}| \rho({\bf q}) |\psi_{S} \rangle ] \,,
\label{eq:A_phi}
\end{equation}
where $\rho({\bf q}) = e^{i{\bf q} \cdot {\bf r}_1} + e^{i{\bf q} \cdot {\bf r}_2}$
\cite{Mahan}.  $A_\phi$ is completely determined
by the charge distribution difference between the two-electron singlet and triplet 
states.  Specifically, in a singlet state (symmetric wave function) the electrons 
tend to distribute themselves across the two dots to minimize their kinetic energy, 
while in a triplet (anti-symmetric wave function) the two electrons avoid each other
to minimize the Coulomb interaction.  For the symmetric double dot we study here, 
the singlet state has larger charge density in between
the two dots, while the triplet has larger charge density at the far ends of the
double dot.  The resulting difference in charge distribution has a finite electrical 
quadrupole moment and gives $A_\phi$ its ${\bf q}$-dependence.

The effective electron-phonon interaction Hamiltonian of Eq.~(\ref{eq:Hamiltonian_eff}) 
is a typical spin-boson Hamiltonian that leads to decay in the off-diagonal element of
the $2 \times 2$ density matrix \cite{Duan}:
\begin{equation}
\rho_{ST} (t) = \rho_{ST} (0) e^{-B^2(t)}\,,
\end{equation}
where the dephasing factor is positive definite:
\begin{equation}
B^2(t) = \frac{2V}{\pi^3 \hbar^2} \int d^3{\bf q} 
\frac{|M({\bf q}) A_\phi({\bf q})|^2}{\omega_{\bf q}^2}  
\sin^2 \frac{\omega_{\bf q} t}{2} \coth \frac{\hbar \omega_{\bf q}}{k_B T} \,.
\label{eq:dephasing_ideal}
\end{equation}
Here $\omega_{\bf q}$ is the angular frequency of the phonons in mode ${\bf q}$.
The derivation of this dephasing formula account for the fact that the
bosonic reservoir is in a thermal equilibrium before getting into contact with the
spin \cite{Duan}.  However, it does not account for the fact that the bosonic modes 
may be dissipative.  Indeed, an ordinary spin-boson calculation of dephasing generally 
assumes a spectral density of the form $1/f^\alpha$, with the implicit assumption
that bosonic modes with a vanishing spectral density at low frequency does not
contribute to dephasing in any significant way.  In the present study, acoustic
phonons do have a vanishing density of state at low frequency.  On the other hand,
we also know that these phonons also decay very fast, with a time scale as short as 
10-100 ps (by anharmonicity induced phonon decay and by irreversibly propagating 
out of the nanostructure).  By adding such a phonon decay channel (for simplicity,
we assume all phonon modes have the same decay rate $\gamma$), we obtain a modified
expression for the dephasing of the off-diagonal density matrix element:
\begin{eqnarray}
\rho_{ST} (t) & = & \rho_{ST} (0) e^{-B_1^2(t)-B^2_2(t)}\,, 
\nonumber \\
B_1^2(t) & = & \frac{V}{\pi^3 \hbar^2} \int d^3{\bf q} 
\frac{|M({\bf q}) A_\phi({\bf q})|^2}{\omega_{\bf q}^2 +(\gamma/2)^2}  
\nonumber \\
& & \times \left\{ \frac{\omega_{\bf q}^2 - (\gamma/2)^2 }{ \omega_{\bf q}^2 
+ (\gamma/2)^2 } \left( 1- e^{-\frac{\gamma}{2}t} \cos \omega_{\bf q} t 
\right) \right. \nonumber \\
& & \left. - \frac{\omega_{\bf q} \gamma/2}{ \omega_{\bf q}^2 + (\gamma/2)^2 }
e^{-\frac{\gamma}{2}t} \sin \omega_{\bf q} t \right\} 
\coth \frac{\hbar \omega_{\bf q}}{k_B T} \,, 
\label{eq:dephasing1} \\
B^2_2(t) & = & \frac{V}{\pi^3 \hbar^2} \int d^3{\bf q} 
\frac{|M({\bf q}) A_\phi({\bf q})|^2}{\omega_{\bf q}^2 + (\gamma/2)^2}  
\left(\frac{\gamma}{2} t \right) \coth \frac{\hbar \omega_{\bf q}}{k_B T} 
\nonumber \\
& = & \Gamma_{ST} t \,.
\label{eq:dephasing2} 
\end{eqnarray}
At the limit that phonon decay rate $\gamma \rightarrow 0$, $B_1^2(t) \rightarrow 
B^2 (t)$ while
$B_2^2(t) \rightarrow 0$.  For a finite $\gamma$, corresponding to a dissipative 
phonon reservoir, we obtain an additional exponential decay of the off-diagonal 
density matrix element in Eq.~(\ref{eq:dephasing2}) compared to the non-dissipative 
reservoir result of 
Eq.~(\ref{eq:dephasing_ideal}).  The rate of this exponential decay 
$\Gamma_{ST}$ is proportional to both the phonon decay rate $\gamma$ and the geometric 
factors that determine the dephasing factor $B_1^2(t)$.

\begin{figure}
\includegraphics[width=2.8in]{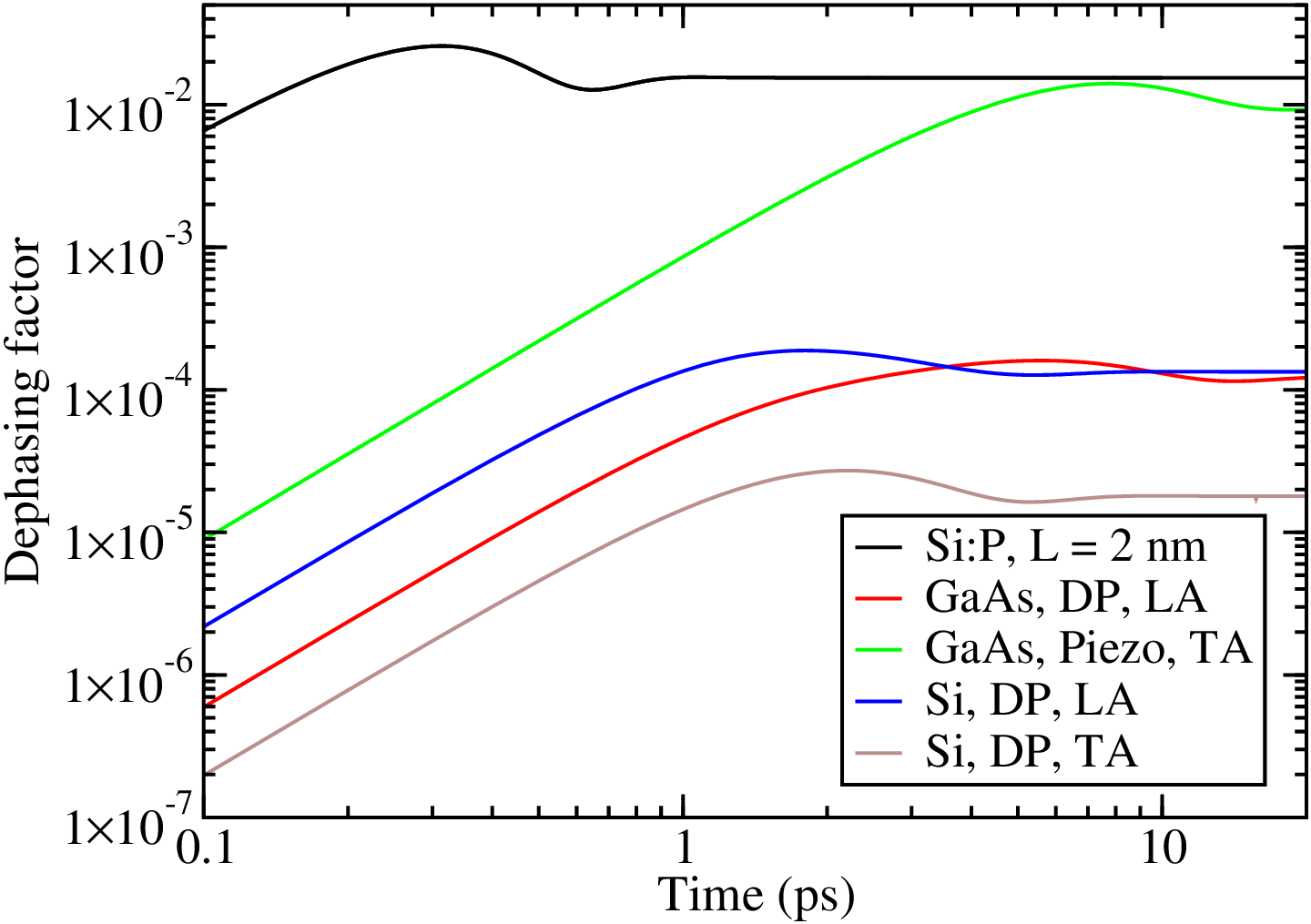}
\caption{
(Color online) Two-spin dephasing in a double quantum dot induced by a 
non-dissipative phonon reservoir.  The solid curve represents $B^2(t)$ for
a P pair in Si separated by 4 nm (The Bohr radius for Si:P is $\lesssim 2$ nm).
All the other curves are for double quantum dots with an interdot separation
of 40 nm and single dot orbital radius of 20 nm.  More specifically, the red
dotted curves are for deformation potential (DP) coupling to longitudinal acoustic
(LA) phonons in GaAs; the green dashed curve is for piezoelectric (PE) coupling
to TA phonons in GaAs; the blue dot-dashed curve is for DP coupling to LA phonons
in Si; and the brown dot-dot-dashed curve is for DP coupling to TA phonons in
Si. 
}
\label{fig:dephasing_nondissipative}
\end{figure}

Let us now examine the dynamical behaviors of the dephasing factors $B^2(t)$ 
and $B^2_2(t)$.
In Fig.~\ref{fig:dephasing_nondissipative} we show the typical behavior of the 
dephasing factor $B^2(t)$ in the absence of phonon decay for various types of
electron-phonon interactions in GaAs and Si.  There are two interesting features 
all the curves in Fig.~\ref{fig:dephasing_nondissipative} share.  At very short 
times ($t \ll 1$ ps), the increase of $B^2(t)$
is quadratic, which originates from Taylor expansion of the $\sin^2 \omega_{\bf q} t
/2$ factor in the integrand at the small $t$ limit.  At long times all the curves 
saturate, which means 
that dephasing does not increase with time anymore, so that it 
corresponds more to a finite loss of contrast than the conventional complete
decay of off-diagonal density matrix elements.  The transition between the 
quadratic increase and the saturation happens between 1 and 10 ps for double dots
with dot separation of about 40 nm and somewhat shorter for P pairs because
this time is essentially determined by the interdot distance divided by the
speed of sound ($\sim 8 \times 10^3$ m/s in Si and $3.7 \times 10^3$ m/s in GaAs).  
Mathematically the long-time saturation
can be understood by writing $2\sin^2 \omega_{\bf q} t /2$ as 
$1 - \cos \omega_{\bf q} t$.  Since acoustic phonon spectrum is continuous,
the cosine term leads to a vanishing contribution to the integral at large times, 
which leaves the dephasing factor determined by a constant integral that is 
independent of time.  Physically, this saturation is due to the fact that 
long-time dephasing is determined by the low-frequency part of the 
spectrum of the bosonic reservoir, while phonon density of state vanishes 
quadratically at low frequency.  In other words, non-dissipative acoustic phonons
simply form an inefficient dephasing reservoir as compared to other charge
fluctuation reservoirs such as fluctuating charge traps, which have a $1/f$
spectral density.

The magnitudes of the saturated dephasing shown in 
Fig.~\ref{fig:dephasing_nondissipative} give a clear sense of the relative 
importance of various types of electron-phonon interactions.  Specifically,
in GaAs the piezoelectric (PE) coupling to transverse acoustic (TA) phonons produces 
the strongest dephasing effect, while in Si the deformation potential (DP) coupling to 
longitudinal acoustic (LA) phonons is the most important.  Here the phosphorus 
dimer has the strongest dephasing because it has one-order-of-magnitude smaller 
inter-donor distance, so that the dominant contribution to its dephasing comes from
higher energy phonons, which have higher density of states, compared to those 
for quantum dots.

\begin{figure}
\includegraphics[width=2.8in]{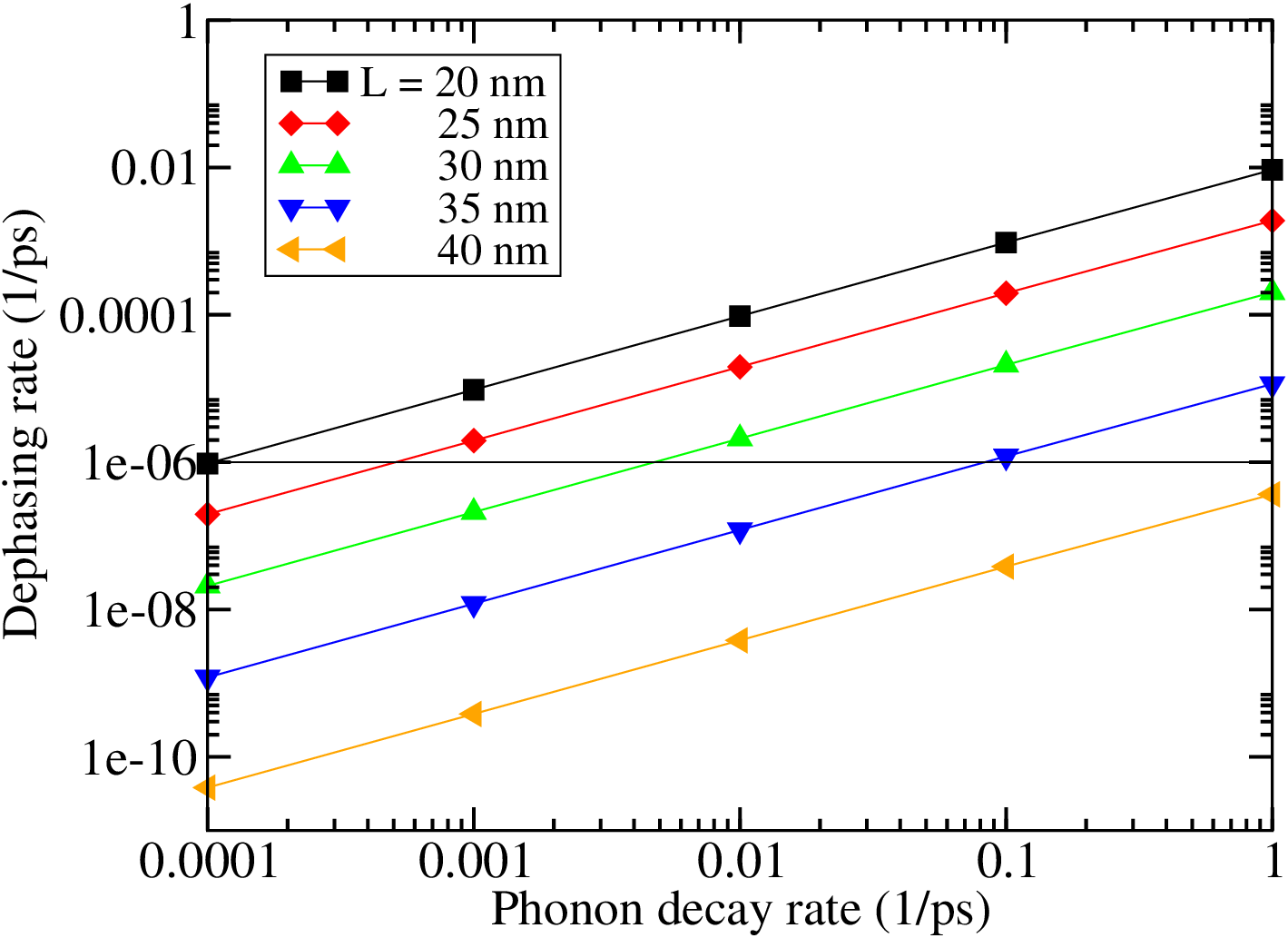}
\caption{
(Color online) Phonon induced two-spin dephasing rate as a function of phonon decay
rate in a GaAs double dot for various interdot separation distance.  The horizontal 
line is drawn at a dephasing time of 1 $\mu$s, approximately the decoherence times 
measured in Refs.~\cite{Petta,Koppens}.  The single dot wave function radius for all 
the data is 20 nm.
}
\label{fig:dephasing_dissipative_g}
\end{figure}

When phonon decay is included, the most important additional effect is the
added exponential dephasing $e^{-\gamma_{ST} t}$.
In Fig.~\ref{fig:dephasing_dissipative_g} we plot the dephasing rate $\Gamma_{ST}$ 
as a function
of the interdot distance and the phonon decay rate $\gamma$.  As an example we plot
the dephasing rate in a GaAs double dot due to PE coupling to TA
phonons.  This is by far the strongest dephasing channel in either GaAs or Si, as
clearly indicated in Fig.~\ref{fig:dephasing_nondissipative}.  Our results show
that for faster phonon decay, we need to keep the interdot distance sufficiently
large in order to obtain longer singlet-triplet dephasing time.

\begin{figure}
\includegraphics[width=2.8in]{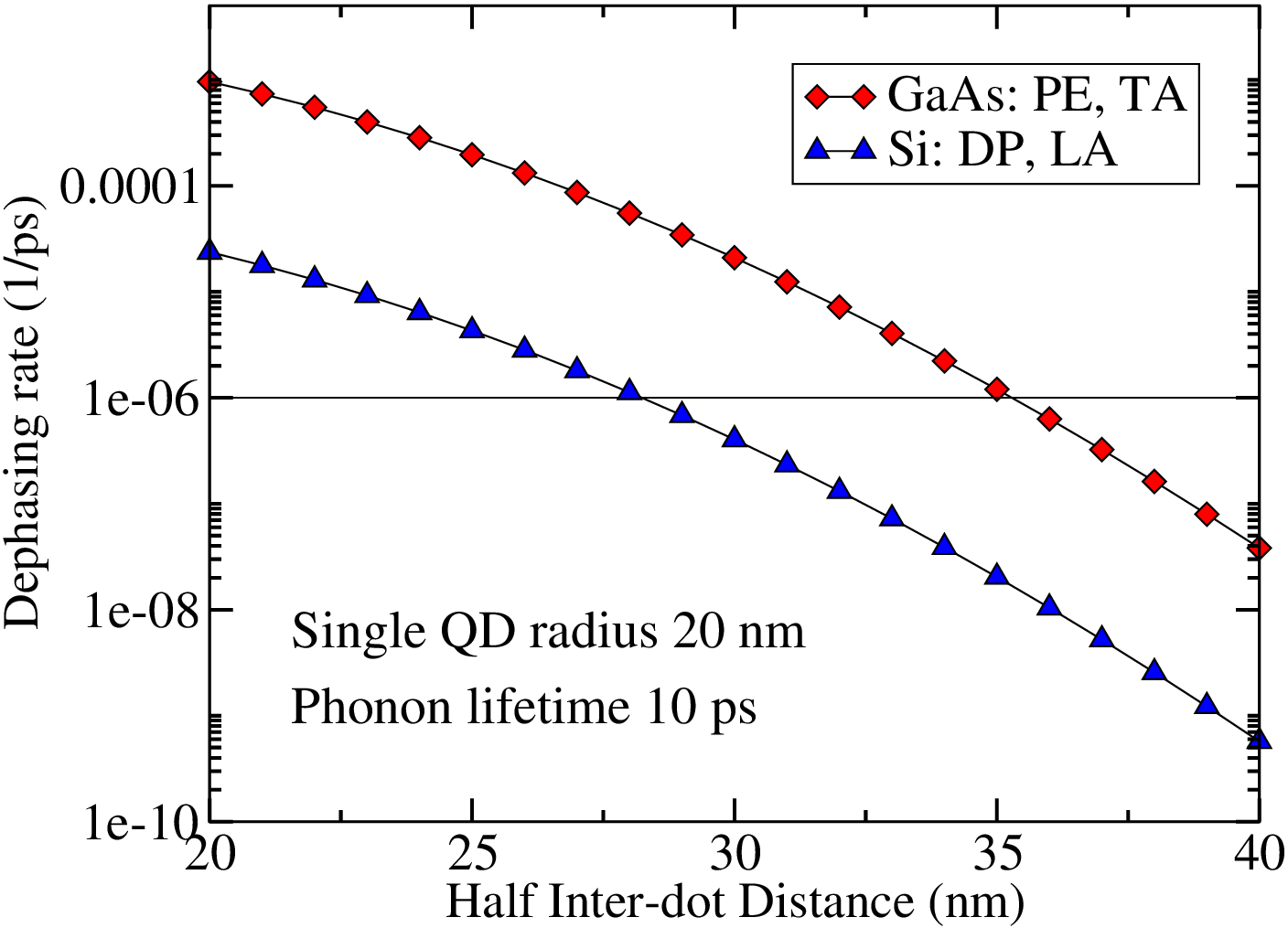}
\caption{
(Color online) Phonon-induced two-spin dephasing rate as a function of the interdot
separation for a GaAs and a Si double quantum dot.  The horizontal line is again 
drawn at a dephasing time of 1 $\mu$s.  For GaAs the curve is for piezoelectric (PE) 
coupling to TA phonons, while for Si it is for deformation potential (DP) coupling 
to LA phonons. 
}
\label{fig:dephasing_dissipative_L}
\end{figure}

In Fig.~\ref{fig:dephasing_dissipative_L} we plot the phonon-induced dephasing rate
between singlet and triplet states in double dots in both GaAs (diamond symbols) and 
Si (triangular symbols) as functions of the interdot distance $L$.  The strong 
dependence on 
$L$ originates from the fact that charge distribution difference between the 
two-electron singlet and triplet states is directly dependent on interdot wave
function overlap: The smaller the overlap, the smaller the difference in charge
distribution, and the smaller the phonon-induced dephasing.  The Si data is about
two orders of magnitude smaller than in GaAs, consistent with what is shown in
Fig.~\ref{fig:dephasing_nondissipative}, and is determined by the fact that
deformation potential interaction in Si is simply a weaker interaction than 
piezoelectric interaction in GaAs.

The magnitude of phonon-induced dephasing to two-spin states in a double dot
depends directly on how fast the phonons themselves relax.  If phonon relaxation
is dominated by internal mechanisms such as phonon anharmonicity \cite{Ridley},
with a time scale much longer than nanoseconds, the phonon-induced dephasing 
would be relatively slow, with a time scale above $\mu$s.  On the other hand,
if phonons escape the nanostructure rapidly, in the order of 10 ps to 1 ns, 
the double dot will need to be well separated for phonon-induced dephasing to 
be sufficiently slow.  Whether this dephasing is slow enough, we need to compare 
it with the speed of gating.

\begin{figure}
\includegraphics[width=2.8in]{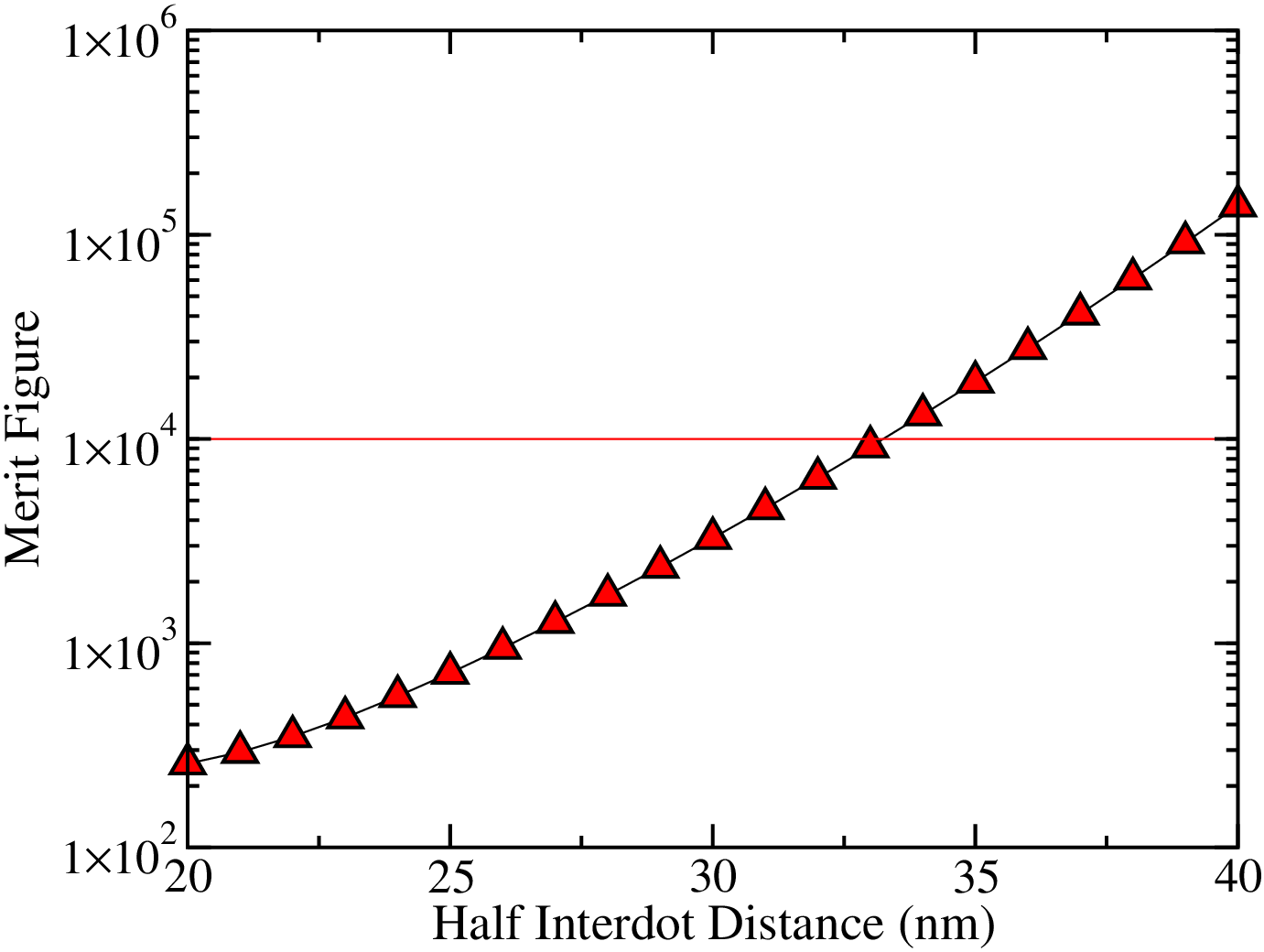}
\caption{
(Color online) Merit figure based on phonon-induced dephasing of two-spin states
in a GaAs double dot as a function of half interdot distance.  
We draw a line at $10^4$ as the nominal threshold for
fault tolerant quantum computation.  Therefore the double dot (with single-dot
wave function radius at 20 nm) should be kept apart further than 70 nm.
}
\label{fig:merit}
\end{figure}

In Fig.~\ref{fig:merit} we plot the two-spin merit figure $\cal{M}$ as a 
function of the interdot distance for double dots in GaAs.  
Here the merit figure is defined as
the ratio between a typical exchange gate time given by $\hbar/J$ ($J$ is
the exchange splitting) and the two-spin decay time given by $1/\gamma_{ST}$:
$\cal{M} = \hbar \gamma_{ST}/J$.  The exchange splitting $J$ is calculated
within the Heitler-London model with a quartic confinement 
potential \cite{BLD}.  The radius of the single dot electron wave function is
20 nm.  We choose a phonon decay time of 10 ps as a worst case
scenario.  The increase of the merit figure at larger inter-dot 
distance reflects the fact that the exchange splitting and the phonon-induced
dephasing have different dependence on the interdot overlap integral $S$: 
$J \sim S^2$, while $\gamma_{ST} \sim S^4$.  The results shown in this figure
reveal that for a two-dot exchange gate to operate with a low error rate, 
slower operation with smaller interdot overlap is preferable.  We do not
have any data for Si quantum dot in this figure.  Calculating exchange 
interaction in a Si double dot requires much more sophisticated quantum
chemical approaches than a simple Heitler-London approximation 
\cite{HD_PRA,Friesen} because in Si the interaction effect is stronger
compared to GaAs (larger effective mass and smaller dielectric constant),
so that Heitler-London approximation does not adequately account for the
two-electron correlation.  For the current evaluation, it is sufficient
to point out that Fig.~\ref{fig:dephasing_dissipative_L} above indicates
that phonon-induced dephasing is about two orders of magnitude weaker in Si 
than in GaAs, while exchange coupling should only be somewhat smaller than in GaAs.  
Therefore overall there should a gain of at least one 
order of magnitude in the merit figure when moving from GaAs to Si.

We have shown here that while the phonon reservoir is treated
as a continuum of states, in the absence of phonon relaxation it does not
lead to complete dephasing of the two-spin states.  This somewhat 
counter-intuitive result can be understood from two different perspectives.
One is the quadratically vanishing phonon density of state at low frequency,
as we mentioned previously.  The other is the fact that the electron-phonon
interaction matrix element vanishes at high phonon frequency for the quantum
dot orbital states.  For example, with a Gaussian envelope function, the
matrix element $\langle L | e^{ikx} | L\rangle \sim e^{k^2a^2/4}$ where $k$
is the phonon wave vector and $a$ is the wave function width.  Thus for large
$k$ these matrix elements vanish rapidly.  Therefore the part of the phonon
spectrum that actually contributes to spin dephasing is relatively small.
For a quantum dot with ground state wave function radius in the order of 20 nm,
only those acoustic phonons with energy below 1 meV are relevant.
In a realistic lattice, the electron states are always dressed by the phonons.
What the results in Fig.~\ref{fig:dephasing_nondissipative} show is that the
dressed states still have their majority spectral weight in the bare two-electron
states.  On the other hand, when phonons can relax completely, they bring the 
electrons into contact with an even larger reservoir that is not closed,
so that the two-spin states can dephase completely.

In our calculations we consider double quantum dots that are symmetric: the two
dots are the same in size and are not voltage biased, in the traditional 
Loss-DiVincenzo configuration for implementing an exchange gate.  In some 
recently studied double dot systems a voltage bias is applied between the
two dots so that the system is close to the bias point where two-dot
singlet state and one of the double-occupied singlet states are degenerate 
\cite{Petta}.
In this configuration the charge distribution difference between the ground 
singlet
and triplet states are more dramatic because the singlet state has a finite
spatial component of doubly occupied state while the triplet does not have
such a component.  In our calculation this should lead to a much larger 
$A_\phi$ defined in Eq.~(\ref{eq:A_phi}), which in turn leads to a larger
spin-boson coupling matrix element.
Accordingly the phonon-induced dephasing should in general
be faster than in the symmetric situation.  A complete calculation for this
situation is beyond the scope of the current paper.

In conclusion, we have studied phonon-induced dephasing between two-electron 
singlet and triplet spin states in a semiconductor double quantum dot.  We find
that this dephasing is important for tightly coupled double dots, especially 
when phonon decay is taken into consideration.  We re-derive the expression
for dephasing in the spin-boson model with a dissipative reservoir, and quantify
the the two-spin dephasing in both GaAs and Si double dots.

We thank the hospitality of the Joint Quantum Institute (JQI) at the 
University of Maryland, where this work is finished.  We thank NSA, LPS, 
ARO, JQI, and DARPA QuEST for financial support.  We also thank useful
discussions with Peter Yu, Sankar Das Sarma, and Susan Coppersmith.

\end{document}